\documentclass[sn-nature]{sn-jnl}
\usepackage{lmodern}
\usepackage{graphicx}%
\usepackage{multirow}%
\usepackage{amsmath,amssymb,amsfonts}%
\usepackage{amsthm}%
\usepackage{mathrsfs}%
\usepackage[title]{appendix}%
\usepackage{xcolor}%
\usepackage{textcomp}%
\usepackage{manyfoot}%
\usepackage{booktabs}%
\usepackage{algorithm}%
\usepackage{algorithmicx}%
\usepackage{algpseudocode}%
\usepackage{listings}%
\usepackage[separate-uncertainty=true,multi-part-units=single]{siunitx}
\usepackage{caption}
\usepackage{tabularx}
\usepackage{bm}
\usepackage{comment}

\usepackage{pdfpages}

\DeclareSIUnit{\atomicunit}{au}

\newcommand{\ket}[1]{\vert #1 \rangle}

\renewcommand{\theta}{\vartheta}

\newcommand{\revFR}[1]{{\color{black}#1}}

\begin{document}

\title{Restoring the Conical Intersection Topology using Convex Density Functional Theory}

\author[1]{\fnm{Federico} \sur{Rossi}}

\author[2]{\fnm{Tommaso} \sur{Giovannini}}

\author*[1]{\fnm{Henrik} \sur{Koch}}
\email{henrik.koch@ntnu.no}

\affil[1]{\orgdiv{Department of Chemistry}, \orgname{Norwegian University of Science and Technology}, \orgaddress{\city{Trondheim}, \postcode{7491}, \country{Norway}}}

\affil[2]{\orgdiv{Department of Physics}, \orgname{University of Rome Tor Vergata}, \orgaddress{\city{Rome}, \postcode{00133}, \country{Italy}}}

\abstract{
Conical intersections are central to the description of photophysics and photochemistry. Nevertheless, in non-adiabatic molecular dynamics simulations, they are fundamentally challenging for single-reference electronic structure methods. Density functional theory (DFT) and its time-dependent extension (TDDFT) represent the most widely used theoretical approaches in physics, chemistry, and biology. However, the treatment of ground and excited states as separate problems leads to breakdowns in the topological structure of potential energy surfaces near conical intersections. In this work, we solve this long-standing issue by presenting Convex DFT, a framework that, by explicitly enforcing convexity of the variational problem within an appropriately defined subspace, guarantees a unique and continuous electronic solution across regions of degeneracies. We demonstrate that \revFR{Convex DFT} yields smooth and physically meaningful intersection seams by comparison with \revFR{multireference wave function methods.} In this way, we establish the method as a robust and computationally efficient DFT approach for treating electronically degenerate regions. These developments represent a critical step toward reliable non-adiabatic simulations beyond the limitations of conventional TDDFT.} 

\maketitle

\newpage

\section*{Introduction}\label{sec:intro} 

Conical intersections govern some of the most fundamental photoinduced processes in nature, providing ultrafast pathways for radiationless energy relaxation and chemical conversion \cite{domcke2011conical,domcke2012role,levine2007isomerization,levine2019conical}. They are central to our understanding of photophysics, photochemistry, and photobiology, from the photostability of biomolecules to charge and energy transfer in complex molecular systems \cite{polli2010conical,garavelli1997relaxation,schapiro2011using,boeije2023one}. From a theoretical perspective, these electronically degenerate regions arguably represent the clearest failure of single-reference electronic-structure methods, most notably Hartree-Fock and Kohn-Sham (KS) density functional theory (DFT) \cite{huix2015description}. This limitation is particularly relevant for DFT, which has emerged as the most widely used electronic structure method due to its favorable balance between accuracy and computational efficiency \cite{matsika2021electronic,dreuw2005single}. Overcoming this failure would therefore unlock the use of DFT in a well-known problematic regime, substantially expanding its applicability across multiple physicochemical phenomena.

The origin of the DFT breakdown in describing ground state conical intersections lies in the fact that the ground and excited states are treated as independent problems: the ground state is obtained variationally through a self-consistent field (SCF) optimization, while excited states are introduced a posteriori via linear-response time-dependent formalisms (TDDFT) \cite{ullrich2012time,hirata1999time}. As a result, the coupled topological structure of electronic states that characterizes a conical intersection is \revFR{problematic from the onset} \cite{gozem2014shape,taylor2023description,kjonstad2025understanding}. This decoupling manifests itself as bifurcations in the energy landscape \cite{cizek1971,cui2013proper}. Near regions of strong non-adiabatic coupling, the SCF optimization problem is no longer well behaved: the energy functional becomes non-convex and admits multiple stationary solutions\cite{capelle2007degenerate}. Standard SCF algorithms may then converge to distinct local minima or saddle points depending on the initial guess\cite{almlof1982principles}, leading to an ambiguous and often unphysical description of the electronic structure near the intersection\cite{filatov2013assessment}. These multiple solutions are not mere numerical artifacts, but rather reflect a deeper inconsistency in the single-reference framework when applied to intrinsically multi-state phenomena \cite{kjonstad2025understanding,burton2022energy}. This becomes even more problematic in the context of non-adiabatic molecular dynamics. When nuclear trajectories evolve around a conical intersection, the presence of discontinuities and bifurcations in the electronic energy surfaces can induce spurious state hopping events between electronic solutions \cite{tapavicza2008mixed}. In many cases, the energy surfaces intersect in a wide region rather than a single point, compromising the reliability of surface-hopping or mixed quantum-classical dynamics simulations. Several strategies have been proposed to address this problem \cite{matsika2021electronic}, spanning from approaches that modify the linear-response or post-SCF treatment to repair the local branching space \cite{li2014configuration,shu2017dual,teh2019simplest}, and formal developments that target the underlying variational structure of the problem \cite{shao2003spin,filatov2015spin,evangelista2013orthogonality,schmerwitz2024saddle,duston2025conical}. Despite these important advances, a KS-DFT framework in which the electronic solution remains unique, continuous, and variationally well defined across regions of degeneracy is still missing.

In this work, we solve this long-standing issue by developing Convex DFT (CVX-DFT), within the Tamm-Dancoff approximation (TDA). In particular, by following the framework recently developed for the Hartree-Fock method \cite{rossi2026convex}, we reformulate the KS-DFT variational problem to explicitly enforce convexity of the energy functional within a suitably defined subspace, overcoming the limitations of previously proposed corrective strategies. Near a degeneracy, the orbital Hessian develops a vanishing or negative eigenvalue associated with the problematic excitation direction. In CVX-DFT the lowest-curvature mode from the orbital optimization is removed by projection, yielding a strictly convex problem that is free of bifurcations and admits a unique, continuous solution. The excluded excitation channel is then included in a second step, through a diagonalization in the full single-excitations space, which gives consistent ground and excited state energies. This two-step procedure provides a smooth and topologically correct description of the electronic structure near degeneracies. 

Embedding the convex optimization into the DFT formalism retains the computational efficiency and favorable treatment of electron correlation offered by modern exchange-correlation functionals, while overcoming the pathological behavior of conventional DFT near conical intersections. The resulting CVX-DFT formulation provides a robust and unified description of electronically degenerate regions of potential energy surfaces, and lays the foundations for a reliable, physically consistent, yet cost-effective description of non-adiabatic phenomena.

\section{Results}
To assess the performance of CVX-DFT in describing the correct topology of ground state conical intersections, we \revFR{first} consider three molecular systems: protonated formaldimine, azobenzene, and the retinal model protonated Schiff base PSB3. These systems are well-established benchmarks of non-adiabatic photochemistry and have previously been studied with both single-reference and multireference methods \cite{taylor2023description,yu2015probing,tuna2015assessment}. For each system, we construct two-dimensional potential energy surface scans, using the branching plane spanned by the gradient difference $\mathbf{g}$ and the derivative coupling $\mathbf{h}$ for formaldimine and PSB3. For azobenzene, we used a linear interpolation of internal coordinates combined with an additional torsional degree of freedom. Where conventional LR-TDDFT/TDA yields discontinuous or unphysical surfaces due to bifurcations, CVX-DFT produces a unique, smoothly varying electronic solution across the intersection region. Throughout, we restrict ourselves to the adiabatic approximation of linear-response TDDFT. Incorporating effects outside this approximation leads to non-linear contributions that allow coupling between ground and excited states \cite{taylor2023description,ullrich2012time}. We compare our results against conventional LR-TDDFT/TDA and, where available, with multireference methods including state average complete active space self-consistent field theory (SA-CASSCF)\cite{roos1980complete} and extended multistate complete active space second-order perturbation theory (XMS-CASPT2)\cite{shiozaki2011communication}. Unless otherwise stated, the BHHLYP functional \cite{becke1993new} is used as the preferred choice, having been shown to yield minimum energy conical intersection geometries in closest agreement with high-level multireference calculations \cite{filatov2013assessment}. All DFT and CVX-DFT calculations are carried out using a locally modified version of eT\cite{folkestad20201,folkestad20262} using the recent TDDFT implementation \cite{marrazzini2021multilevel,giovannini2024time}, while all CASSCF and XMS-CASPT2 results are obtained using BAGEL\cite{shiozaki2018bagel}. Further information on geometries and computational details is provided in the Supplementary Notes 2-6. \revFR{The results are complemented by a broader set of conical intersections and Franck--Condon geometries.}

\subsection{Protonated formaldimine}
\begin{figure}[!htb]
    \centering
    \includegraphics[width=\textwidth]{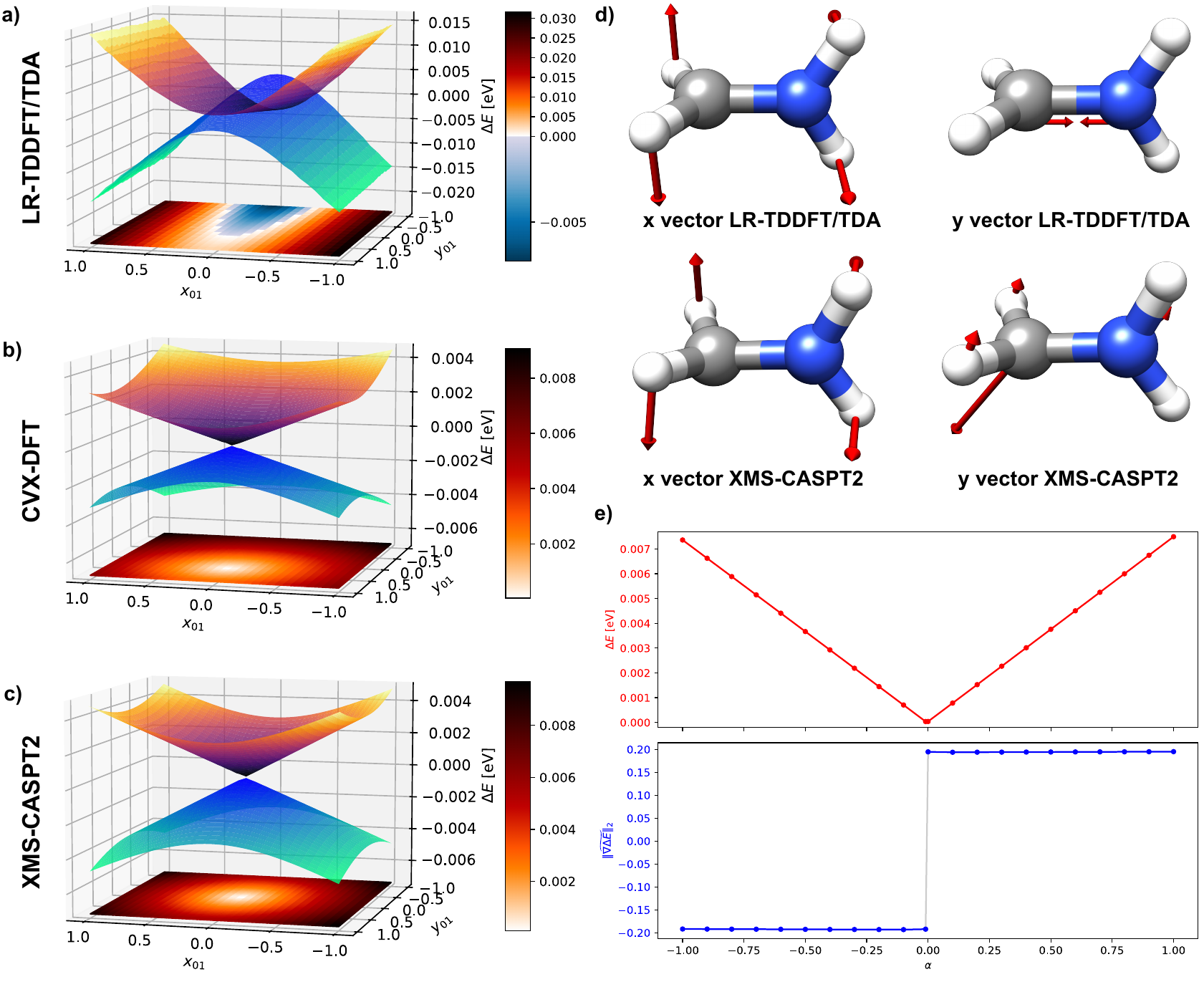}
    \caption{\footnotesize \textbf{S$_0$/S$_1$ conical intersection in formaldimine.} Potential energy surfaces computed with (a) LR-TDDFT/TDA and (b) CVX-DFT, using cc-pVDZ basis set and PBE0 functional, and (c) XMS-CASPT2/cc-pVTZ. All displacement steps are scaled by a factor of $10^{-3}$. For each point the energies are plotted in eV relative to the average energy of the states. The colormap at the bottom of each plot indicates the energy gap $E_1 - E_0$. \revFR{In (d),} the geometry at the conical intersection and the displacement vectors $\mathbf{x}$ and $\mathbf{y}$, the orthogonalized $\mathbf{g}$ and $\mathbf{h}$, are graphically depicted for LR-TDDFT/TDA (top) and XMS-CASPT2 (bottom) \cite{taylor2023description}. \revFR{Panel (e) shows the energy gap $\Delta E = E_1 - E_0$ (upper) and the signed gradient norm $\widetilde{\|\nabla E\|}_2$ (lower) along a linear path through the CVX-DFT intersection, computed with central differences using a displacement of $10^{-7}$.}}
    \label{fig:for_complete}
\end{figure}

As a first system, we consider protonated formaldimine, the simplest model for the chromophore in rhodopsin, the protonated Schiff base of retinal. \revFR{The} ground state conical intersection has previously been studied by Taylor et al.\cite{taylor2023description}, and here we adopted the same geometry, branching plane vectors, and functional (PBE0 \cite{adamo1999toward}). The potential energy surfaces computed with LR-TDDFT/TDA produce a linear intersection (Fig. \ref{fig:for_complete}a) and exhibit a region where the ordering of the states is interchanged. In this region, the excitation energy becomes negative, a direct manifestation of the pathological behavior described above. This is consistent with the findings of the original work, but we note that extending the branching plane reveals the intersection as part of a larger closed intersection ring \cite{taylor2023description}. The CVX-DFT framework recovers a proper conical topology (Fig. \ref{fig:for_complete}b), demonstrating that the convexity constraint successfully restores the qualitatively correct intersection topology. Notably, the potential energy surfaces obtained with XMS(3)-CASPT2(6,4), using three-state averaging and active space (6,4)\cite{barbatti2006ultrafast}, yield a similar conical topology (Fig. \ref{fig:for_complete}c).

\revFR{
A necessary condition for the reliable use of any electronic structure method in non-adiabatic molecular dynamics is the availability of smooth, well-defined energy gradients in the vicinity of conical intersections. Discontinuities or divergences in the gradient, which arise naturally from the bifurcating surfaces of conventional LR-TDDFT/TDA, would lead to unphysical forces on the nuclei and compromise simulations. While analytical gradients for CVX-DFT are not yet implemented, we assess their behavior numerically along a linear path through the formaldimine conical intersection, parameterized as $\mathbf{R}(\alpha) = \mathbf{R}_0 + \alpha\,(\mathbf{g} + \mathbf{h})$, where $\mathbf{R}_0$ is the geometry at the CVX-DFT intersection and $\alpha$ is a displacement parameter. The results are shown in Fig.~\ref{fig:for_complete}e, for $\alpha \in [-1,1]\times10^{-3}$. The upper panel shows the energy gap $\Delta E = E_1 - E_0$, which decreases to $1.6\times 10^{-3}$ eV at $\alpha = 0$. The lower panel shows the signed norm of the numerical gradient, defined as $\widetilde{\|\nabla E\|}_2 = \|\nabla E\|_2 \cdot \mathrm{sgn}\!\left(\sum_i [\nabla E]_i\right)$, which is approximately constant on both sides of the intersection and changes sign at the minimum of the energy gap. A point was included for $\alpha = -0.01 \times 10^{-3}$ to show the behavior of the numerical gradient immediately before the crossing. This profile is the expected signature of a well-behaved crossing between two adiabatic surfaces and confirms that CVX-DFT yields continuous, physically meaningful gradients across the intersection region. The implementation of analytical gradients and non-adiabatic couplings, which would make CVX-DFT directly applicable to non-adiabatic dynamics simulations, is left for future work.\\\\
}

\clearpage
\subsection{Azobenzene}
\begin{figure}[!htb]
    \centering
    \includegraphics[width=\textwidth]{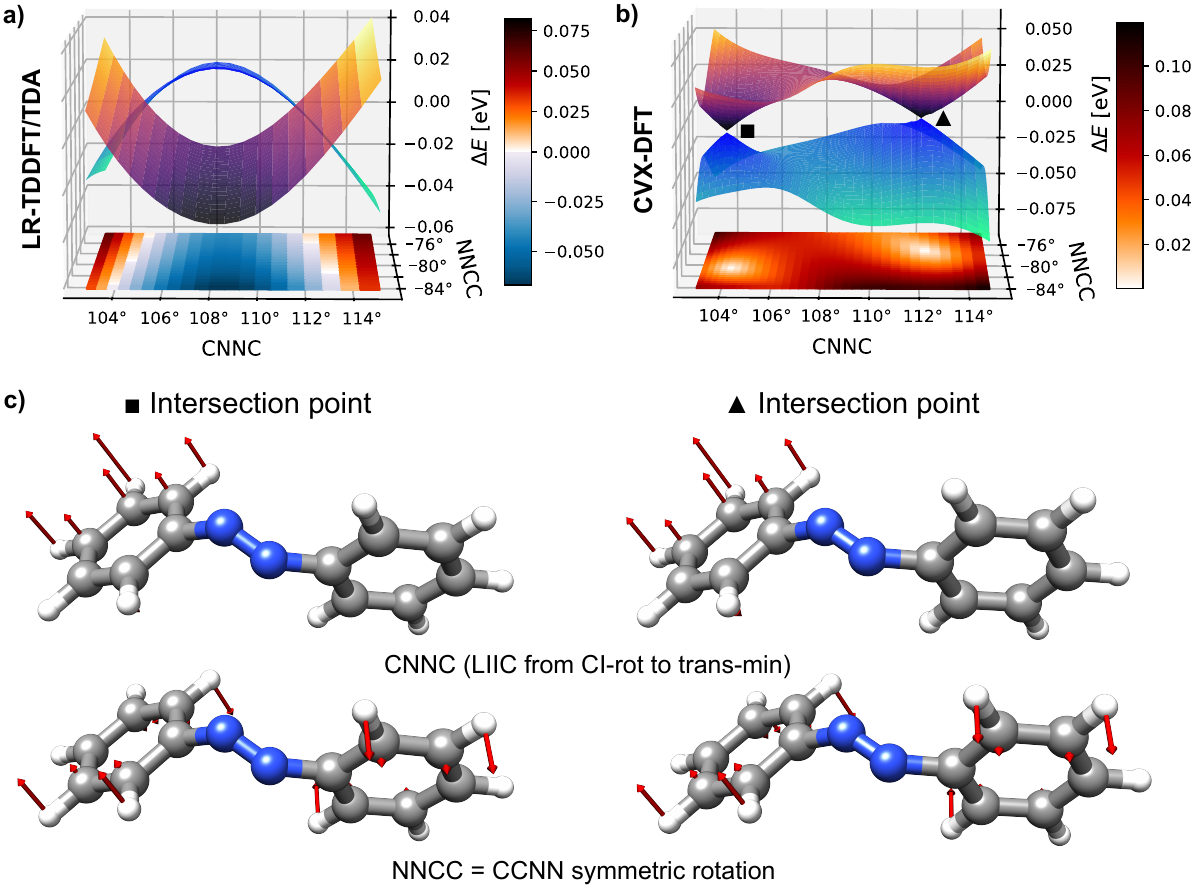}
    \caption{\footnotesize \textbf{S$_0$/S$_1$ conical intersection in azobenzene.} Potential energy surfaces computed with (a) LR-TDDFT/TDA and (b) CVX-DFT, using 6-31G basis set and BHHLYP functional. Vectors describing the two coordinates that define the plane are illustrated in (c) for both intersection points identified by CVX-DFT. For each point, the energies are plotted in eV relative to the average energy of the states. The colormap at the bottom of each plot indicates the energy gap $E_1 - E_0$.}
    \label{fig:azo_combined}
\end{figure}

Another well-known problematic system is azobenzene. In Fig. \ref{fig:azo_combined}, we compare the ground and first excited state potential energy surfaces for LR-TDDFT/TDA and CVX-DFT. The scan is constructed from a linear interpolation of internal coordinates (LIIC) connecting the S$_0$/S$_1$ rotational conical intersection to the trans minimum on S$_0$, both optimized at SA5-CASSCF(6,6) level in a previous work \cite{yu2015probing}. The CNNC torsional angle serves as a measure of displacement along this path. A second coordinate is introduced as an additional symmetric rotation of both phenyl rings around the CCNN and NNCC dihedral angles, applied on top of the LIIC geometry. With LR-TDDFT/TDA (Fig. \ref{fig:azo_combined}a), the two surfaces cross each other on two lines, with the in-between region characterized by negative excitation energies. CVX-DFT instead finds two distinct crossing points (Fig. \ref{fig:azo_combined}b), each lying close to the corresponding linear intersections found by LR-TDDFT/TDA. The geometries at the two crossing points are similar but distinct, as shown in Fig. \revFR{S13} of Supplementary Note \revFR{6}.

\clearpage
\subsection{Retinal model PSB3}
\begin{figure}[!htb]
    \centering
    \includegraphics[width=\textwidth]{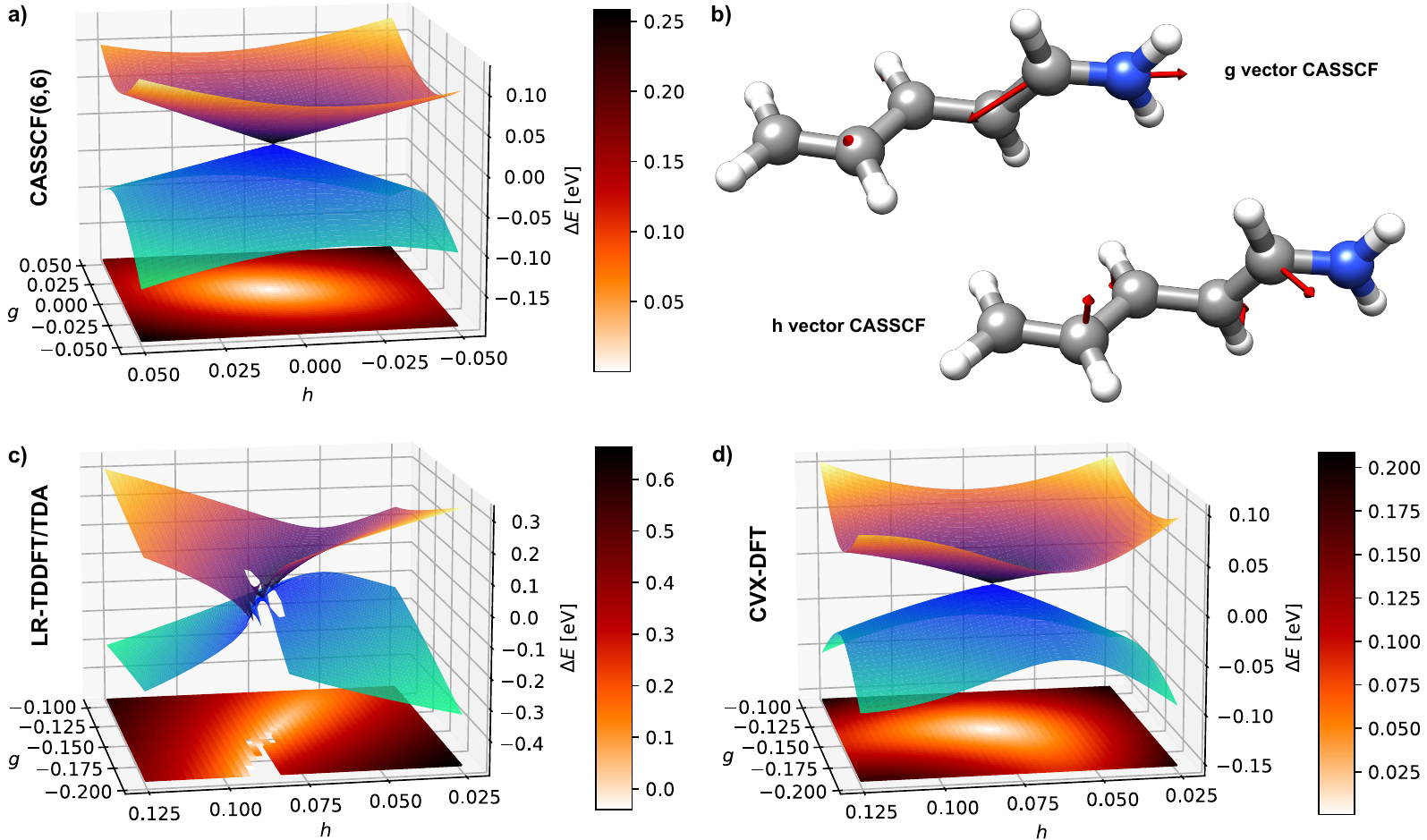}
    \caption{\footnotesize \textbf{S$_0$/S$_1$ conical intersection in retinal model PSB3.} The minimum energy conical intersection (MECI) geometry, optimized at the SA2-CASSCF(6,6)/cc-pVDZ level, is shown in (b) alongside the corresponding orthogonalized branching plane vectors $\mathbf{g}$ and $\mathbf{h}$. Potential energy surfaces computed at SA2-CASSCF(6,6) using these vectors are shown in (a). Potential energy surfaces computed with (c) LR-TDDFT/TDA and (d) CVX-DFT, using cc-pVDZ basis set and BHHLYP functional, are obtained with the same geometry and branching plane vectors as in (b). For each point the energies are plotted in eV relative to the average energy of the states. The colormap at the bottom of each plot indicates the energy gap $E_1 - E_0$.}
    \label{fig:ret_combined}
\end{figure}

Finally, we consider PSB3, another widely studied model system for retinal. The S$_0$/S$_1$ MECI and orthonormalized $\mathbf{g}$ and $\mathbf{h}$ (see Fig. \ref{fig:ret_combined}b) were obtained with BAGEL at SA2-CASSCF(6,6)/cc-pVDZ level, starting from the SA2-CASSCF(6,6)/6-31G* MECI optimized by Tuna et al. \cite{tuna2015assessment}. The resulting CASSCF potential energy surfaces (Fig. \ref{fig:ret_combined}a) display a conical intersection between S$_0$ and S$_1$, centered at the origin (0,0). For both LR-TDDFT/TDA and CVX-DFT, the branching plane is described using the CASSCF vectors. For this reason, the intersection point is \revFR{located} at (-0.1465, 0.0750) Bohr. With LR-TDDFT/TDA (Fig. \ref{fig:ret_combined}c), convergence issues appear in the region close to the intersection, leaving a gap where the method could not converge. This stands the well-known issues associated with common TDDFT: the surfaces on either side of this region are discontinuous and the conical shape is missing. CVX-DFT (Fig. \ref{fig:ret_combined}d) restores the conical shape, avoiding convergence issues. Furthermore, the potential energy surfaces are smooth and continuous throughout the branching plane, in agreement with the reference CASSCF(6,6) results.

\revFR{
\subsection{Benchmarking}
Assessing the generality of the CVX-DFT topology beyond the three systems discussed above, we apply the method to a broader set of molecular systems with known ground state conical intersections, drawn from the benchmark study of Nikiforov et al.\cite{nikiforov2014assessment} and later Lee et al. \cite{lee2019conical}, which established a set of reference geometries and branching plane vectors with MRSF-TD-DFT/6-31+G**, SF-TD-DFT/6-31+G** and  MRCISD/6-31+G** methods. The molecules in the set cover a range of chromophore types and conical intersection topologies. For each system, we determine an approximated CVX-DFT crossing point moving along the MRSF-TD-DFT branching plane vectors $\mathbf{g}$ and $\mathbf{h}$ of Lee et al., with the 6-31+G** basis set and BHHLYP functional as in the original work. All potential energy surface scans around each intersection are provided in Supplementary Note 4, and all show an intersection with a conical topology. To provide a quantitative assessment of the intersection geometry, we compute the root-mean-square deviation (RMSD) of each CVX-DFT approximate MECI with respect to the MRCISD reference geometry of the original work. As reported in Table S5 of Supplementary Note~4, CVX-DFT locates the intersection in geometry space with an accuracy comparable to that of MR-SF-TDDFT across all systems in the set. We note that the comparison is inherently approximate: the branching plane vectors used to locate the CVX-DFT intersection are those of MR-SF-TDDFT, not of CVX-DFT itself, so the re-optimized geometry is not a true MECI for CVX-DFT.

Until now, we have only considered geometries close to ground state degeneracies. At Franck--Condon (FC) geometries, the ground and excited states are well separated and the orbital Hessian is positive definite. The behavior of CVX-DFT at such non-degenerate geometries depends on the symmetry of the projected states: when all projected states are not totally symmetric, the coupling terms vanish by symmetry and CVX-DFT reduces to standard LR-TDDFT/TDA exactly. Instead, when one or more projected states are totally symmetric, a non-zero coupling may arise even at non-degenerate geometries, introducing a small but finite difference with respect to LR-TDDFT/TDA. A systematic benchmark of vertical excitation energies against aug-cc-pVTZ theoretical best estimates from the QUEST databases \cite{loos2018mountaineering,loos2020mountaineering,veril2021questdb} confirms that these differences remain small across all tested systems, establishing that CVX-DFT can serve as a drop-in replacement for LR-TDDFT/TDA at Franck--Condon geometries. Full details and the complete comparison table are provided in Supplementary Note~5.

}

\section{Discussion}
In this work, we have shown that the pathological behavior of conventional KS-DFT near ground state conical intersections can be traced back to a loss of convexity in the underlying orbital optimization problem, and that this issue can be resolved within a unified DFT framework. In fact, in CVX-DFT, ground and excited states emerge consistently from a single diagonalization, naturally restoring the topology required for a correct description of conical intersections. Across three representative benchmark systems, this formulation converts the discontinuous, bifurcating, or ill-defined topologies obtained with conventional TDDFT into smooth and physically meaningful intersection seams, in qualitative agreement with multireference reference methods. \revFR{The robustness of the approach is further supported by the extended benchmark, in which CVX-DFT recovers correct conical intersection geometries across a chemically diverse set of systems. Furthermore, a benchmark of vertical excitation energies at Franck--Condon geometries confirms that CVX-DFT introduces negligible changes relative to LR-TDDFT/TDA in the non-degenerate regime, establishing that the method can be used across both degenerate and non-degenerate regions of the potential energy surface.} These results \revFR{show} that the failure  at conical intersections of standard linear-response TDDFT within the adiabatic approximation is not an unavoidable limitation of the KS-DFT method itself, but rather a consequence of the non-convex structure of its conventional formulation. \revFR{We expect an extension of the method beyond the Tamm-Dancoff approximation to be possible using the CVX-DFT ground state as the reference state for the full time-dependent DFT treatment, which will be explored in future work.}
\\\\
The CVX-DFT framework paves the way for applying DFT to electronically degenerate regimes that have long remained outside its reliable domain of applicability. The method retains the favorable computational scaling and broad functional flexibility of modern DFT, while removing the bifurcations and discontinuities that compromise non-adiabatic simulations. \revFR{A timing analysis (Supplementary Note~3) shows that the total wall time of CVX-DFT at a conical intersection geometry is approximately four times that of a DIIS-based LR-TDDFT/TDA that converges. Using a Newton-Raphson approach for LR-TDDFT/TDA is 3.5 times slower than CVX-DFT. This suggests an adaptive procedure where CVX-DFT is used when the states are close in energy. Such an approach was successfully used for coupled cluster dynamics \cite{angelico2026spectroscopic}.  Unlike spin-flip approaches, which address the S$_0$/S$_1$ intersection problem by starting from a high-spin reference state, CVX-DFT operates entirely within the closed-shell singlet framework. This avoids the introduction of bifurcations or new intersection problems in the high-spin manifold, where the reference state can itself become degenerate with other states in different regions of the potential energy surface.} In this way, \revFR{CVX-DFT} provides a promising foundation for the treatment of photochemical relaxation, radiationless decay, and excited-state reactivity in systems that are far beyond the scope of state-of-the-art multireference approaches. Our work thereby lays the groundwork for accurate, routine, yet first-principle large-scale simulations of non-adiabatic phenomena of complex molecular systems, materials, and hybrid systems.

\section{Methods}
\subsection{Convex Density Functional Theory}
We consider the KS-DFT framework, in which, for a global hybrid functional, the total energy $\mathcal{E}$ can be expressed as
\begin{equation}
\mathcal{E}[\mathbf{D}] = \text{Tr}\,\mathbf{h}\mathbf{D}+\frac{1}{2}\text{Tr}\,\mathbf{D}\mathbf{J}(\mathbf{D})-\frac{1}{2}c_\text{x}\text{Tr}\,\mathbf{D}\mathbf{K}(\mathbf{D})+ E_\text{xc}[\mathbf{D}],
\end{equation}
where $\mathbf{D}$ is the density matrix and $\mathbf{h}$, $\mathbf{J}(\mathbf{D})$, and $\mathbf{K}(\mathbf{D})$ are the integral matrices for the one-electron, Coulomb, and exact-exchange parts of the Hamiltonian, respectively. The coefficient $c_\text{x}$ scales the exact-exchange contribution and is equal to 1 for Hartree-Fock and equal to 0 for pure DFT functionals. The density is constructed from a single Slater determinant of KS orbitals, which are obtained by solving the KS equations with an effective local potential that accounts for electron correlation contributions, resulting in an additional contribution $E_\text{xc}[\mathbf{D}]$.\\
To describe variations of the KS determinant, we employ an exponential orbital-rotation parameterization. The KS wave function $\Psi^\text{KS}$ is written as an exponential of an anti-Hermitian single-excitation operator applied to an initial determinant $\Phi_0$ constructed from orbitals obtained from the superposition of atomic densities (SAD) \cite{van2006starting},
\begin{align}
    \ket{\Psi^\text{KS}} = \exp(\sum_{ai} \kappa_{ai}E^-_{ai}) \ket{\Phi_0} \label{eq:Slater}
\end{align}
where $E^-_{ai}=E_{ai}- E_{ia}$ generates occupied-virtual orbital rotations parameterized by $\kappa_{ai}$. This parameterization preserves orthonormality of the orbitals and spans the space of the Slater determinants. In terms of density matrices, this transformation can be expressed as\cite{coriani2010atomic}
\begin{equation}
\mathbf{D(\bm{\kappa})} = \text{exp}(-\bm{\kappa})\mathbf{D}\,\text{exp}(\bm{\kappa}).
\end{equation}
Following the convex Hartree-Fock formulation \cite{rossi2026convex}, we introduce an additional set of orbital-rotation parameters $\bm{\gamma}$, used exclusively to analyze the local curvature of the energy functional. The energy is therefore considered a function of the density $\mathbf{D}(\bm{\gamma})$ induced by these rotations,
\begin{equation}
\mathcal{E}[\mathbf{D}(\bm{\gamma})] = \text{Tr}\,\mathbf{h}\mathbf{D}(\bm{\gamma})+\frac{1}{2}\text{Tr}\,\mathbf{D}(\bm{\gamma})\mathbf{J}(\mathbf{D}(\bm{\gamma}))-\frac{1}{2}c_\text{x}\text{Tr}\,\mathbf{D}(\bm{\gamma})\mathbf{K}(\mathbf{D}(\bm{\gamma}))+ E_\text{xc}[\mathbf{D}(\bm{\gamma})],
\end{equation}
The first and second derivatives of the energy with respect to $\bm{\gamma}$ define the gradient $\mathbf{G}^{(0)}$ and the Hessian $\mathbf{G}^{(1)}$ in the orbital-rotation space \cite{larsen2000hartree}
\begin{align}
   G^{(0)}_{ai}(\mathbf{0}) &= \frac{\partial \mathcal{E}}{\partial \gamma_{ai}}\Big|_{\bm\gamma=\mathbf{0}},\\ 
   G^{(1)}_{ai,bj}(\mathbf{0}) &= \frac{\partial^2 \mathcal{E}}{\partial \gamma_{ai}\partial \gamma_{bj}}\Big|_{\bm\gamma=\mathbf{0}}.
   \label{eq:Hessian_HF}
\end{align}
In the presence of degeneracies or near-degeneracies in the KS reference, the orbital Hessian generally exhibits zero or negative eigenvalues. These correspond to directions in orbital-rotation space along which the energy surface is locally flat or concave, leading to multiple stationary points and ill-conditioned optimization behavior. In particular, when the KS reference corresponds to a ground state that is degenerate with respect to a single excitation, the Hessian exhibits a zero eigenvalue associated with that excitation direction, with nearby regions of negative curvature. This renders the orbital optimization problem non-convex. To restore convexity, we explicitly project out the problematic orbital-rotation directions. Specifically, diagonalizing the orbital Hessian yields eigenvectors $\mathbf{r}_n$ and eigenvalues $\lambda_n$. The lowest-curvature mode $\mathbf{r}_1$, associated with a vanishing or negative eigenvalue, is removed from the optimization problem. This is achieved by redefining the orbital-rotation gradient $\tilde{\mathbf{G}}^{(0)}$ through projection,
\begin{align}
\tilde{\mathbf{G}}^{(0)} = \mathbf{G}^{(0)} - \mathbf{r}_1 (\mathbf{r}_1^T \mathbf{G}^{(0)}).
\end{align}
The same projection is applied to the orbital-rotation parameters themselves to ensure consistency of the reduced parameter space 
\begin{align}
    \tilde{\kappa}_{ai} = \kappa_{ai} - r_{ai,1} \sum_{bj}r_{bj,1} \kappa_{bj}.
\end{align}
The resulting projected Hessian is positive definite by construction, and the corresponding orbital-optimization problem is strictly convex. Standard second-order optimization methods can therefore be applied without encountering saddle points or flat directions.
\\
The projection procedure removes physically relevant excitation directions from the variational optimization. To recover these contributions, we perform a final diagonalization in the space consisting of the optimized KS determinant, the projected excitation direction, and the remaining orthogonal single-excitation manifold.
This leads to a CIS-like generalized eigenvalue problem, yielding corrected ground and excited state energies that incorporate the previously excluded excitation channel
\begin{equation}
    \mathbf{H}^{\mathrm{FS}}\mathbf{x}_n = \mathcal{E}_n \mathbf{x}_n.
     \label{eq:FSeigen}
\end{equation}
Multiple states can be added to the projection operator, which can be helpful in situations where different states intersect the ground state in different regions of the \revFR{potential energy surfaces}. All the equations and the generalization of the method to multiple projected states can be found in Supplementary Note 1 of the Supplementary Information. \revFR{We note that since the eigenvectors of Eq. \ref{eq:FSeigen} are orthogonal, they lead to well-defined transition moments among the states.}

\clearpage

\section*{Data availability}
Geometries xyz files and all computational details can be found in Supplementary Notes 2-6. 

\section*{Code availability}
The code used in this study will be made publicly available at \url{https://etprogram.org/}. Prior to its public release, access can be granted from the corresponding author upon reasonable request.

\bibliography{sn-bibliography}

\section*{Acknowledgments}
We thank Jack T. Taylor and Basile F. Curchod for helpful correspondence that ensured the correct reproduction of the formaldimine results. This work was supported by the European Research Council (ERC) under the European Union's Horizon 2020 Research and Innovation Program (grant agreement No. 101020016). This work has also received funding from the ERC under the European Union’s Horizon Europe research and innovation programme (grant no. 101219149, project CHOPIN).

\section*{Author contributions}
F.R., T.G., and H.K. conceived the CVX-DFT framework, analyzed the data and wrote the paper. F.R. developed the implementation in eT and performed all the calculations. H.K. supervised the project.

\section*{Competing interests}
The authors declare no competing interests.




\section*{Supplementary material}

# Contents

# Supplementary Note 1
# Hamiltonian matrix in the new basis

## 1.1 Single projected state

For Hartree-Fock, the full space Hamiltonian matrix with a single projected state [1] can be expressed in terms of $\hat{H}^{\text{HF}}$ as

$$\mathbf{H}^{\text{FS}} = \begin{pmatrix} \langle \text{HF}|\hat{H}^{\text{HF}}|\text{HF}\rangle & \langle \text{HF}|\hat{H}^{\text{HF}}|\text{R}_1\rangle & 0 \\ \langle \text{R}_1|\hat{H}^{\text{HF}}|\text{HF}\rangle & \langle \text{R}_1|\hat{H}^{\text{HF}}|\text{R}_1\rangle & \langle \text{R}_1|\hat{H}^{\text{HF}}|\tilde{\nu}\rangle \\ 0 & \langle \tilde{\mu}|\hat{H}^{\text{HF}}|\text{R}_1\rangle & \langle \tilde{\mu}|\hat{H}^{\text{HF}}|\tilde{\nu}\rangle \end{pmatrix}, \quad (1)$$

where $\langle \tilde{\mu}|\hat{H}^{\text{HF}}|\text{HF}\rangle = \langle \text{HF}|\hat{H}^{\text{HF}}|\tilde{\nu}\rangle = 0$ due to the previous optimization of the KS determinant. Here, $\mu$ and $\nu$ are limited to single excitations and their corresponding projected versions are defined as

$$\langle \tilde{\mu}| = \langle \mu| - \langle \mu|\text{R}_1\rangle\langle \text{R}_1| \quad (2)$$

$$|\tilde{\nu}\rangle = |\nu\rangle - |\text{R}_1\rangle\langle \text{R}_1|\nu\rangle \quad (3)$$

$$|\text{R}_1\rangle = \frac{1}{\sqrt{2}} \sum_{ai} r_{ai,1} E_{ai}|\text{HF}\rangle \quad (4)$$

For DFT, we can express Eq. 5 in terms of $E_0$, $\mathbf{G}^{(0)}$ and $\mathbf{A}$, which already contain the exchange-correlation contributions, as

$$\mathbf{H}^{\text{FS}} = \begin{pmatrix} E_0 & \mathbf{G}^{(0)}\mathbf{r}_1 & \mathbf{0} \\ \mathbf{G}^{(0)}\mathbf{r}_1 & E_0 + \mathbf{r}_1^T \mathbf{A} \mathbf{r}_1 & V_\nu \\ \mathbf{0} & V_\mu & Z_{\mu\nu} + \delta_{\mu\nu} E_0 \end{pmatrix}, \quad (5)$$

where

$$E_0 = \text{Tr}\,\mathbf{hD} + \frac{1}{2}\text{Tr}\,\mathbf{DJ}(\mathbf{D}) - \frac{1}{2}c_{\text{x}}\text{Tr}\,\mathbf{DK}(\mathbf{D}) + E_{\text{xc}}[\rho] \quad (6)$$

$$G^{(0)}_{ai}(\mathbf{0}) = F_{ai} = h_{ai} + J_{ai} - c_{\text{x}} K_{ai} + F^{\text{xc}}_{ai} \quad (7)$$

$$G^{(1)}_{ai,bj}(\mathbf{0}) = A_{ai,bj} + B_{ai,bj} \quad (8)$$

$$A_{ai,bj} = (\delta_{ij} F_{ab} - \delta_{ab} F_{ij}) + (ai|bj) - c_x(ab|ij) + f^{\text{xc}}_{ai,bj} \quad (9)$$

$$B_{ai,bj} = (ai|bj) - c_x(aj|bi) + f^{\text{xc}}_{ai,jb} \quad (10)$$

$$V_\mu = A_{\mu\nu} r_{1,\nu} - r_{1,\mu}(\mathbf{r}_1^T \mathbf{A} \mathbf{r}_1) \quad (11)$$

$$Z_{\mu\nu} = A_{\mu\nu} - r_{1,\mu} V_\nu - V_\mu r_{1,\nu} - r_{1,\mu} r_{1,\nu}(E_0 + \mathbf{r}_1^T \mathbf{A} \mathbf{r}_1) \quad (12)$$

$$F^{\text{xc}}_{ai} = \frac{\partial E_{\text{xc}}[\rho]}{\partial D_{ai}} = \int \frac{\partial E_{\text{xc}}[\rho]}{\partial \rho(\mathbf{r})} \frac{\partial \rho(\mathbf{r})}{\partial D_{ai}} d\mathbf{r} \quad (13)$$

$$f_{ai,bj}^{\text{xc}} = \frac{\partial^2 E_{\text{xc}}[\rho]}{\partial D_{ai} \partial D_{bj}} \tag{14}$$

Above, we used the convention of labeling occupied MOs with $i,j$, virtual MOs with $a,b$ and general MOs with $p,q,r,s$, and indicated with $(pq|rs)$ the two-electron integrals. Note that in Eq. 9 the Fock terms are not diagonal, since the ground state problem is only solved in the projected subspace.

## 1.2 Multiple projected states

Following the formalism above, for Hartree-Fock the full space Hamiltonian matrix with $N$ projected states can be expressed in terms of $\hat{H}^{\text{HF}}$ as

$$\mathbf{H}^{\text{FS}} = \begin{pmatrix} \langle \text{HF}|\hat{H}^{\text{HF}}|\text{HF}\rangle & \langle \text{HF}|\hat{H}^{\text{HF}}|\text{R}_1\rangle & \ldots & \langle \text{HF}|\hat{H}^{\text{HF}}|\text{R}_N\rangle & 0 \\ \langle \text{R}_1|\hat{H}^{\text{HF}}|\text{HF}\rangle & \langle \text{R}_1|\hat{H}^{\text{HF}}|\text{R}_1\rangle & \ldots & \langle \text{R}_1|\hat{H}^{\text{HF}}|\text{R}_N\rangle & \langle \text{R}_1|\hat{H}^{\text{HF}}|\tilde{\nu}\rangle \\ \vdots & \vdots & \ddots & \vdots & \vdots \\ \langle \text{R}_N|\hat{H}^{\text{HF}}|\text{HF}\rangle & \langle \text{R}_N|\hat{H}^{\text{HF}}|\text{R}_1\rangle & \ldots & \langle \text{R}_N|\hat{H}^{\text{HF}}|\text{R}_N\rangle & \langle \text{R}_N|\hat{H}^{\text{HF}}|\tilde{\nu}\rangle \\ 0 & \langle \tilde{\mu}|\hat{H}^{\text{HF}}|\text{R}_1\rangle & \ldots & \langle \tilde{\mu}|\hat{H}^{\text{HF}}|\text{R}_N\rangle & \langle \tilde{\mu}|\hat{H}^{\text{HF}}|\tilde{\nu}\rangle \end{pmatrix} \tag{15}$$

where now the projected single excitations are defined as

$$\langle \tilde{\mu}| = \langle \mu| - \sum_I \langle \mu|\text{R}_I\rangle \langle \text{R}_I| \tag{16}$$

$$|\tilde{\nu}\rangle = |\nu\rangle - \sum_I |\text{R}_I\rangle \langle \text{R}_I|\nu\rangle \tag{17}$$

Again, for DFT Eq. 15 can be expressed as

$$\mathbf{H}^{\text{FS}} = \begin{pmatrix} E_0 & \mathbf{G}^{(0)}\mathbf{r}_1 & \ldots & \mathbf{G}^{(0)}\mathbf{r}_N & \mathbf{0} \\ \mathbf{G}^{(0)}\mathbf{r}_1 & E_0 + \mathbf{r}_1^T \mathbf{A} \mathbf{r}_1 & \ldots & \mathbf{r}_1^T \mathbf{A} \mathbf{r}_N & V_{1,\nu} \\ \vdots & \vdots & \ddots & \vdots & \vdots \\ \mathbf{G}^{(0)}\mathbf{r}_N & \mathbf{r}_N^T \mathbf{A} \mathbf{r}_1 & \ldots & E_0 + \mathbf{r}_N^T \mathbf{A} \mathbf{r}_N & V_{N,\nu} \\ \mathbf{0} & V_{1,\mu} & \ldots & V_{N,\mu} & Z_{\mu\nu} + \delta_{\mu\nu} E_0 \end{pmatrix} \tag{18}$$

where

$$W_{IJ} = \mathbf{r}_I^T \mathbf{A} \mathbf{r}_J \tag{19}$$

$$V_{I,\mu} = A_{\mu\nu} r_{I,\nu} - \sum_J r_{J,\mu}(E_0 \delta_{IJ} + W_{JI}) \tag{20}$$

$$Z_{\mu\nu} = A_{\mu\nu} - \sum_I r_{I,\mu} V_{I,\nu} - \sum_I V_{I,\mu} r_{I,\nu} - \sum_{IJ} r_{I,\mu}(E_0 \delta_{IJ} + W_{IJ}) r_{J,\nu} \tag{21}$$

# Supplementary Note 2
# Computational details

CVX-DFT equations are solved following the algorithm proposed in Ref. 1. All calculations are performed using a 30th-order quadrature, with the radial threshold set to $10^{-6}$. The DFT functionals are implemented in eT through an interface with the LibXC library[2, 3]. All 2D scans are run with the initial geometry, **g** and **h** vectors reported in the next section. These vectors are obtained in Hartree/Bohr, then orthonormalized and used as displacement vectors in the same unit of the initial geometry $\mathbf{r}_0$, defining a new geometry as $\mathbf{r}_0 + \alpha\mathbf{g} + \beta\mathbf{h}$. Geometries will be identified with the corresponding pair of coefficients $(\alpha, \beta)$.
Conversion factors: 1Å = 1.8897259886 Bohr, 1 Hartree = 27.2114079527 eV

## 2.1 Formaldimine

All LR-TDDFT/TDA and CVX-DFT calculations are run with a locally modified version of eT[4, 5], with basis set cc-pVDZ and functional PBE0, to be consistent with the original optimization [6] as reported in the geometry section below. For the potential energy surfaces plot in Fig. 1a-b, a grid of 33x33 equally spaced points centered at the geometry in Tab. S3 was created for $x \in [-0.001, +0.001]$Å and $y \in [-0.001, +0.001]$Å. CVX-DFT calculations only include the lowest excited state in the projection operator.

All XMS(3)-CASPT2(6,4) calculations are run with BAGEL[7], with basis set cc-pVTZ, auxiliary basis set cc-pVTZ-JKFit and "cartesian" : "true". The list of relevant keywords used is reported in Tab. S1 and is consistent with a previous study [6]. For the potential energy surfaces plot in Fig. 1c, a grid of 33x33 equally spaced points centered at the geometry in Tab. S6 was created for $x \in [-0.001, +0.001]$Å and $y \in [-0.001, +0.001]$Å.

**Table S1**: Parameters used for BAGEL calculations in protonated formaldimine.
CASSCF: "nstate" : 3, "nact" : 4, "nclosed" : 5, "charge": 1, "active" : [5,7,8,9]
CASPT2: "ms" : "true", "xms" : "true", "sssr" : "true", "shift" : 0.3, "frozen" : true

## 2.2 Azobenzene

Both LR-TDDFT/TDA and CVX-DFT 2d plots are run with a locally modified version of eT[4, 5], with basis set 6-31G and functional BHHLYP. A set of 1601 geometries was obtained as a Linear Interpolation of Internal Coordinates (LIIC) connecting the $S_0/S_1$ rotational conical intersection to the trans minimum on $S_0$, both optimized at SA5-CASSCF(6,6) level in a previous work [8]. For CVX-DFT, a set of 34 LIIC geometries corresponding to equally spaced indices from 262 to 493 was chosen as one of the scan coordinates, using the value of the CNNC torsional angle as a measure

of displacement along this path. The second coordinate corresponds to an additional symmetric rotation of each phenyl ring around the CCNN and NNCC dihedral angles, choosing 31 equally distant values in $[-84.86°, -74.66°]$. CVX-DFT calculations only include the lowest excited state in the projection operator. For LR-TDDFT/TDA, the grid was created with 21 LIIC geometries corresponding to equally spaced indices between 257 and 597 and then 21 equally distant additional symmetric phenyl rings rotation in $[-85°, -75°]$.

## 2.3 Retinal model PSB3

The SA2-CASSCF(6,6) MECI geometry optimized by Tuna et al. [9] with 6-31G* was optimized again at SA2-CASSCF(6,6) level using BAGEL[7] and cc-pVDZ basis set. The auxiliary basis set cc-pVDZ-JKF was used, with all additional keywords reported in Tab. S2. The orthonormalized **g** and **h** vectors and all the CASSCF calculations on this system were obtained with BAGEL at the same level of theory. For the potential energy surfaces in Fig. 3a, a grid of 41x41 equally spaced points centered at the geometry in Tab. S13 was created for $\mathbf{g} \in [-0.05, +0.05]$ Bohr and $\mathbf{h} \in [-0.05, +0.05]$ Bohr.

Both LR-TDDFT/TDA and CVX-DFT 2d plots are run with a locally modified version of eT [4, 5], with basis set cc-pVDZ and functional BHHLYP. For the potential energy surfaces in Fig. 3c-d, a grid of 41x41 equally spaced points centered at the geometry in Tab. S13 was created for $\mathbf{g} \in [-0.1965, -0.0965]$ Bohr and $\mathbf{h} \in [0.025, 0.125]$ Bohr. Note that the **g** and **h** vectors are obtained at SA2-CASSCF(6,6) level, which results in a grid with the same spacing but with an intersection shifted to the geometry reported in Tab. S16. CVX-DFT calculations only include the lowest excited state in the projection operator.

**Table S2**: Parameters used for BAGEL calculations in retinal model PSB3.
OPTIMIZATION: "opttype" : "meci", "target" : 0, "target2" : 1, "maxstep" : 0.03
CASSCF: "nstate" : 2, "nclosed" : 19, "nact" : 6, "charge": 1

# Supplementary Note 3
# Geometries

## 3.1 Formaldimine

All geometries and displacement vectors for protonated formaldimine were taken from the work of Taylor et al. [6] and are reported below. A detailed description of the optimization procedures can be found in the Supporting Information of that study. All calculations were run in Angstrom.

Table S3: Protonated formaldimine $S_0/S_1$ MECP LR-TDDFT/TDA with cc-pVDZ geometry in Angstrom.

| Atom | $x$ | $y$ | $z$ |
| --- | --- | --- | --- |
| N | 0.7082475627 | 0.0271471256 | 0.0173424147 |
| C | -0.6582583385 | 0.0703534499 | 0.0452205507 |
| H | 1.2813749747 | 0.7436610140 | 0.4774549282 |
| H | 1.2102122950 | -0.7271732350 | -0.4676682844 |
| H | -1.2072815119 | -0.4286168927 | 0.8606727970 |
| H | -1.2070538845 | 0.6047335150 | -0.7474568786 |

Table S4: Protonated formaldimine $S_0/S_1$ MECP LR-TDDFT/TDA orthonormalized g vector.

| Atom | $x$ | $y$ | $z$ |
| --- | --- | --- | --- |
| N | 0.0002274398 | -0.0008371628 | 0.0011387178 |
| C | -0.0004018034 | 0.0015408535 | -0.0023320862 |
| H | -0.0001449954 | 0.2806342076 | -0.4367782423 |
| H | 0.0001521724 | -0.2811110004 | 0.4373756972 |
| H | -0.0177951555 | -0.4033493634 | -0.2588355323 |
| H | 0.0179595815 | 0.4031228553 | 0.2594310189 |

Table S5: Protonated formaldimine $S_0/S_1$ MECP LR-TDDFT/TDA orthonormalized h vector.

| Atom | $x$ | $y$ | $z$ |
| --- | --- | --- | --- |
| N | -0.5433304081 | 0.0207796638 | 0.0139513033 |
| C | 0.8167051782 | -0.0258232995 | -0.0170316156 |
| H | -0.0128528926 | -0.0141368264 | -0.0093169943 |
| H | -0.0134154852 | 0.0126624535 | 0.0080929617 |
| H | -0.1235249574 | -0.0235203286 | 0.0430670689 |
| H | -0.1235817638 | 0.0300384157 | -0.0387627823 |

**Table S6**: Protonated formaldimine $S_0/S_1$ MECX XMS(3)-CASPT2(6,4) with cc-pVTZ geometry in Angstrom.

| Atom | $x$ | $y$ | $z$ |
|---|---|---|---|
| N | 0.703946 | -0.000003 | -0.000013 |
| C | -0.706230 | -0.058493 | -0.037646 |
| H | 1.211952 | 0.750250 | 0.482224 |
| H | 1.277082 | -0.714460 | -0.458992 |
| H | -1.179659 | -0.353299 | 0.896283 |
| H | -1.179742 | 0.668488 | -0.693817 |

**Table S7**: Protonated formaldimine $S_0/S_1$ MECX XMS(3)-CASPT2(6,4) orthonormalized g vector.

| Atom | $x$ | $y$ | $z$ |
|---|---|---|---|
| N | 0.0002172659 | -0.0260069301 | 0.0409335452 |
| C | -0.0005376617 | 0.1128391122 | -0.1767430392 |
| H | 0.0001357967 | 0.2944040343 | -0.4580652016 |
| H | -0.0000296571 | -0.2595289952 | 0.4038576072 |
| H | -0.0691096147 | -0.4331442461 | -0.1442414169 |
| H | 0.0693238715 | 0.3114370255 | 0.3342585048 |

**Table S8**: Protonated formaldimine $S_0/S_1$ MECX XMS(3)-CASPT2(6,4) orthonormalized h vector.

| Atom | $x$ | $y$ | $z$ |
|---|---|---|---|
| N | 0.2528701388 | 0.2120046656 | 0.1363731591 |
| C | -0.5823831219 | -0.5419621332 | -0.3481714125 |
| H | 0.0824745874 | 0.0148402569 | 0.0101211728 |
| H | 0.0314393232 | -0.0288445279 | -0.0191348261 |
| H | 0.1078852723 | 0.1907138674 | 0.0820488744 |
| H | 0.1077137998 | 0.1532478707 | 0.1387630328 |

## 3.2 Azobenzene

The rotational $S_0/S_1$ CI geometry and $S_0$ trans minimum were optimized at SA5-CASSCF(6,6) level with 6-31G in a previous work [8]. Note that the xyz coordinates reported below are constructed from the internal coordinates and differ to what is reported in the original work by a rigid rotation.

Table S9: Azobenzene $S_0/S_1$ rotational CI geometry at SA5-CASSCF(6,6) with 6-31G in Angstrom.

| Atom | $x$ | $y$ | $z$ |
| --- | --- | --- | --- |
| N | 0.00000000 | 0.00000000 | 0.00000000 |
| C | 1.35643181 | 0.00000000 | 0.00000000 |
| C | 2.05954630 | -1.22236297 | 0.00000000 |
| C | 3.43304942 | -1.22165444 | -0.01712061 |
| C | 4.15246952 | -0.02506072 | -0.01756756 |
| C | 3.45765728 | 1.18371817 | -0.00622774 |
| C | 2.08276339 | 1.21261296 | -0.00719077 |
| H | 1.50417662 | -2.13847412 | 0.01480438 |
| H | 3.95744303 | -2.15794175 | -0.02115730 |
| H | 5.22431849 | -0.03546763 | -0.02244168 |
| H | 3.99944403 | 2.11023760 | 0.00073092 |
| H | 1.55069443 | 2.14330902 | -0.00109472 |
| N | -1.05957625 | 0.61405716 | 0.15725201 |
| C | -1.77565612 | 1.14868562 | -0.93392185 |
| C | -1.36672396 | 1.03197404 | -2.26878539 |
| C | -2.12695588 | 1.59868154 | -3.26673084 |
| C | -3.30554506 | 2.28273800 | -2.95899600 |
| C | -3.71535718 | 2.38681461 | -1.62800651 |
| C | -2.96222072 | 1.82129335 | -0.62466757 |
| H | -0.46875348 | 0.49620605 | -2.50461076 |
| H | -1.81293129 | 1.51078181 | -4.28891466 |
| H | -3.89463483 | 2.71993802 | -3.74122860 |
| H | -4.62168069 | 2.90739421 | -1.38674119 |
| H | -3.26020224 | 1.88472605 | 0.40287322 |

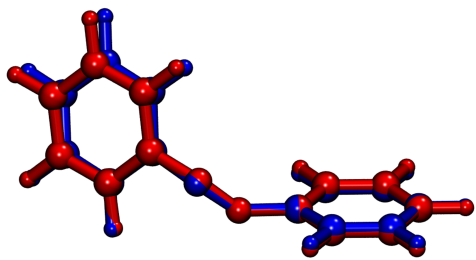

**Fig. S1**: Comparison of the $S_0/S_1$ crossing point ▲ (red) and ■ (blue) of azobenzene calculated at the CVX-DFT/6-31G level.

**Table S10**: Azobenzene $S_0$ trans minimum at SA5-CASSCF(6,6) with 6-31G geometry in Angstrom.

| Atom | $x$ | $y$ | $z$ |
|------|-----|-----|-----|
| N | 0.00000000 | 0.00000000 | 0.00000000 |
| C | 1.41678116 | 0.00000000 | 0.00000000 |
| C | 2.02105807 | -1.25775097 | 0.00000000 |
| C | 3.41614108 | -1.35799128 | -0.00000283 |
| C | 4.19037609 | -0.20063304 | -0.00000318 |
| C | 3.56769709 | 1.05958699 | 0.00000153 |
| C | 2.19829816 | 1.16221257 | 0.00000217 |
| H | 1.39806392 | -2.12979391 | -0.00000152 |
| H | 3.88490150 | -2.32280625 | -0.00000325 |
| H | 5.26116889 | -0.26729129 | -0.00000211 |
| H | 4.16956453 | 1.94782978 | 0.00000660 |
| H | 1.71670517 | 2.11824554 | 0.00000628 |
| N | -0.57983460 | 1.08329421 | -0.00003806 |
| C | -2.00403513 | 1.07623770 | -0.00003391 |
| C | -2.61088686 | 2.32210279 | -0.00012232 |
| C | -3.99486281 | 2.41914839 | -0.00012661 |
| C | -4.76686890 | 1.26744564 | -0.00004162 |
| C | -4.15420090 | 0.01860445 | 0.00004976 |
| C | -2.77395401 | -0.08217992 | 0.00005307 |
| H | -1.99127047 | 3.19656377 | -0.00018926 |
| H | -4.46398635 | 3.38361202 | -0.00019920 |
| H | -5.83747527 | 1.33844213 | -0.00004442 |
| H | -4.75286033 | -0.87158140 | 0.00011871 |
| H | -2.29037667 | -1.03704881 | 0.00012511 |

**Table S11**: Azobenzene $S_0/S_1$ crossing point ■ for CVX-DFT with 6-31G geometry in Angstrom. The geometry corresponds to CNNC angle of 104.1069° (corresponding to geometry 290 of 1601 in the LIIC) and additional symmetric CCNN=NNCC rotation of -81.8000°.

| Atom | $x$ | $y$ | $z$ |
|---|---|---|---|
| N | 0.00000000 | 0.00000000 | 0.00000000 |
| C | 1.36733241 | 0.00000000 | 0.00000000 |
| C | 2.05272164 | -0.17533865 | 1.21676596 |
| C | 3.43067982 | -0.19173460 | 1.23184390 |
| C | 4.16013656 | -0.02233748 | 0.05426005 |
| C | 3.47804311 | 0.16067538 | -1.15032561 |
| C | 2.10384755 | 0.16588016 | -1.19222504 |
| H | 1.48494286 | -0.29301646 | 2.11769942 |
| H | 3.94521185 | -0.32935013 | 2.16332916 |
| H | 5.23201101 | -0.02911080 | 0.07393455 |
| H | 4.03079237 | 0.29763002 | -2.06008451 |
| H | 1.58072875 | 0.30422432 | -2.11742792 |
| N | -0.97881202 | 0.70369612 | 0.12270748 |
| C | -1.93035281 | 1.25745423 | -0.76285540 |
| C | -1.93996457 | 2.61462231 | -1.09907432 |
| C | -2.90033382 | 3.09868249 | -1.96148232 |
| C | -3.85807966 | 2.24162192 | -2.50422434 |
| C | -3.83913839 | 0.88601856 | -2.17329290 |
| C | -2.88144913 | 0.39477424 | -1.31357992 |
| H | -1.19547049 | 3.26637755 | -0.68700697 |
| H | -2.90851660 | 4.14027394 | -2.21833521 |
| H | -4.60218833 | 2.62221474 | -3.17631383 |
| H | -4.57172400 | 0.22313959 | -2.59116533 |
| H | -2.84713666 | -0.64323370 | -1.04993081 |

Table S12: Azobenzene $S_0/S_1$ crossing point ▲ for CVX-DFT with 6-31G geometry in Angstrom. The geometry corresponds to CNNC angle of 112.1384° (corresponding to geometry 401 of 1601 in the LIIC) and additional symmetric CCNN=NNCC rotation of -78.4000°.

| Atom | $x$ | $y$ | $z$ |
|---|---|---|---|
| N | 0.00000000 | 0.00000000 | 0.00000000 |
| C | 1.37340506 | 0.00000000 | 0.00000000 |
| C | 2.04889323 | -0.24795069 | 1.20792160 |
| C | 3.42922817 | -0.26556587 | 1.23220148 |
| C | 4.16425071 | -0.02733747 | 0.07054767 |
| C | 3.48930286 | 0.22675347 | -1.12647045 |
| C | 2.11555248 | 0.23600822 | -1.17523921 |
| H | 1.47423767 | -0.41948421 | 2.09576608 |
| H | 3.93823080 | -0.45838843 | 2.15683057 |
| H | 5.23609900 | -0.03587965 | 0.09539079 |
| H | 4.04813978 | 0.41602362 | -2.02300503 |
| H | 1.59745224 | 0.42889679 | -2.09330513 |
| N | -0.93265585 | 0.75278629 | 0.10467765 |
| C | -1.99612479 | 1.29574850 | -0.65324939 |
| C | -2.12869813 | 2.67282107 | -0.84526872 |
| C | -3.18578276 | 3.16277752 | -1.58410445 |
| C | -4.11812382 | 2.29004025 | -2.14313482 |
| C | -3.97741426 | 0.91486991 | -1.95579419 |
| C | -2.92288883 | 0.41840046 | -1.22009548 |
| H | -1.40252929 | 3.33632822 | -0.41956708 |
| H | -3.28819864 | 4.22056308 | -1.73050213 |
| H | -4.93775764 | 2.67425028 | -2.71841553 |
| H | -4.69139952 | 0.24038072 | -2.38710652 |
| H | -2.79572841 | -0.63447224 | -1.06795323 |

## 3.3 Retinal model PSB3

The $S_0/S_1$ MECI geometry was optimized at SA2-CASSCF(6,6) level with cc-pVDZ using BAGEL, starting from the geometry optimized at the same level with 6-31G* and reported in a previous work [9]. Gradient and non-adiabatic coupling were calculated at the same level using BAGEL. All calculations were run in Bohr.

Table S13: Retinal model PSB3 $S_0/S_1$ MECI SA2-CASSCF(6,6) with cc-pVDZ in Angstrom.

| Atom | $x$ | $y$ | $z$ |
|------|-----|-----|-----|
| C | -2.85180658 | -0.86118147 | 0.80187922 |
| C | -1.47614734 | -0.91110458 | 0.85232569 |
| C | -0.70108048 | -0.06156713 | 0.05319607 |
| C | 0.75703568 | -0.06796911 | 0.02959530 |
| C | 1.51655214 | 0.74918692 | 0.83350444 |
| N | 2.84307092 | 0.79730607 | 0.83912166 |
| H | 1.23057894 | -0.73075303 | -0.68909938 |
| H | -1.23002595 | 0.61440865 | -0.61544478 |
| H | 1.03360223 | 1.41813100 | 1.53441399 |
| H | -0.98138143 | -1.60769710 | 1.51620606 |
| H | 3.32996373 | 1.41655193 | 1.45398540 |
| H | 3.39549598 | 0.22735413 | 0.22946131 |
| H | -3.45888930 | -1.50807843 | 1.42019585 |
| H | -3.36606843 | -0.17335900 | 0.14295194 |

Table S14: Retinal model PSB3 $S_0/S_1$ MECI SA2-CASSCF(6,6) orthonormalized g vector.

| Atom | $x$ | $y$ | $z$ |
|------|-----|-----|-----|
| C | 0.15295686 | 0.05255293 | -0.04093244 |
| C | -0.10503492 | 0.10208898 | 0.21598014 |
| C | 0.01854380 | -0.00974750 | -0.06617427 |
| C | 0.21061970 | 0.20490932 | 0.27199048 |
| C | -0.51821389 | -0.41374262 | -0.05248225 |
| N | 0.43375004 | 0.05902536 | 0.01153129 |
| H | -0.06637253 | 0.13966959 | -0.16295607 |
| H | -0.04222726 | -0.14043709 | -0.13440114 |
| H | -0.00193451 | -0.00096074 | 0.00090501 |
| H | 0.00750266 | 0.01234311 | -0.00685501 |
| H | -0.03686020 | -0.03418602 | -0.06762357 |
| H | -0.03847880 | 0.05015726 | 0.04906085 |
| H | -0.00718166 | -0.02460397 | -0.00962686 |
| H | -0.00706930 | 0.00293138 | -0.00841617 |

**Table S15**: Retinal model PSB3 $S_0/S_1$ MECI SA2-CASSCF(6,6) orthonormalized h vector.

| Atom | $x$ | $y$ | $z$ |
| --- | --- | --- | --- |
| C | -0.06711268 | -0.01839545 | 0.03356709 |
| C | 0.03216736 | 0.33201123 | 0.30631681 |
| C | -0.01156200 | 0.03650071 | -0.02730259 |
| C | -0.10191915 | -0.14761135 | -0.09922452 |
| C | 0.27429289 | -0.23570120 | 0.46830899 |
| N | -0.22188296 | 0.01772707 | -0.05415667 |
| H | 0.02455178 | 0.30896466 | -0.27361342 |
| H | 0.03159811 | -0.28143534 | -0.28757717 |
| H | 0.00098535 | -0.00352961 | 0.00341422 |
| H | -0.00555201 | -0.00478416 | 0.00715692 |
| H | 0.01721808 | 0.05972342 | -0.00684541 |
| H | 0.01976601 | -0.02387689 | -0.02634473 |
| H | 0.00362429 | -0.03011790 | -0.03933002 |
| H | 0.00382486 | -0.00947499 | -0.00436960 |

**Table S16**: Retinal model PSB3 $S_0/S_1$ CVX-DFT intersection point geometry in Angstrom, using BHHLYP and cc-pVDZ basis set. The geometry corresponds to coefficients pair (-0.1465, 0.0750) Bohr of the 2D scan.

| Atom | $x$ | $y$ | $z$ |
| --- | --- | --- | --- |
| C | -2.86632806 | -0.86598569 | 0.80638470 |
| C | -1.46672790 | -0.90584201 | 0.84773913 |
| C | -0.70297695 | -0.05936281 | 0.05724260 |
| C | 0.73666250 | -0.08971304 | 0.00457133 |
| C | 1.56761262 | 0.77190752 | 0.85615948 |
| N | 2.80063852 | 0.79343372 | 0.83607831 |
| H | 1.23669885 | -0.72931856 | -0.68732554 |
| H | -1.22549823 | 0.61412628 | -0.61643885 |
| H | 1.03379131 | 1.41806539 | 1.53447934 |
| H | -0.98218342 | -1.60884387 | 1.51702154 |
| H | 3.33350465 | 1.42157251 | 1.45895620 |
| H | 3.39926351 | 0.22251808 | 0.22461232 |
| H | -3.45818870 | -1.50736635 | 1.41938122 |
| H | -3.36536859 | -0.17396230 | 0.14343098 |

# Supplementary References



\end{document}